\documentclass[conference,compsoc]{IEEEtran}
\usepackage{siunitx}
\usepackage{tabularx}
\usepackage{makecell}
\usepackage{threeparttable}
\usepackage[linesnumbered,ruled,vlined]{algorithm2e}
\usepackage{adjustbox}
\usepackage{booktabs}
\usepackage{cuted}
\usepackage{arydshln}
\usepackage{multirow}  
\usepackage{comment}
\usepackage{caption}
\usepackage{subcaption}
\usepackage{hyperref}  
\usepackage{amsmath}
\usepackage{mathptmx}
\usepackage{xcolor}
\usepackage{tcolorbox}
\usepackage{lipsum}
\usepackage{textcomp}
\usepackage{stfloats}
\usepackage{url}
\usepackage{xurl} 
\usepackage{verbatim}
\usepackage{graphicx}
\usepackage[utf8]{inputenc}  
\usepackage{longtable}  
\usepackage{path}
\usepackage{supertabular}
\usepackage[nocompress]{cite}
\usepackage{pifont}
\ifCLASSINFOpdf
\else
\fi

\usepackage{enumitem}

\begin{document}
%
\title{From IOCs to Regex: Automating CTI Operationalization for SOC with LLMs}

\author{
\IEEEauthorblockN{
Pei-Yu Tseng\IEEEauthorrefmark{1},
Lan Zhang\IEEEauthorrefmark{2},
ZihDwo Yeh\IEEEauthorrefmark{1},
Xiaoyan Sun\IEEEauthorrefmark{3},
Xushu Dai\IEEEauthorrefmark{1},
Peng Liu\IEEEauthorrefmark{1}
}

\IEEEauthorblockA{\IEEEauthorrefmark{1}The Pennsylvania State University, PA, USA}
\IEEEauthorblockA{\IEEEauthorrefmark{2}Northern Arizona University, AZ, USA}
\IEEEauthorblockA{\IEEEauthorrefmark{3}Worcester Polytechnic Institute, MA, USA}

\IEEEauthorblockA{
\IEEEauthorrefmark{1}jerry950909@gmail.com, doyleyeh@gmail.com, xfd5059@psu.edu, pxl20@psu.edu\\
\IEEEauthorrefmark{2}Lan.Zhang@nau.edu\\
\IEEEauthorrefmark{3}xsun7@wpi.edu
}
}
\maketitle

\begin{abstract}
Cyber Threat Intelligence (CTI) reports provide Indicators of Compromise (IOCs) that are essential for security operations. A common way to operationalize IOCs is through \textbf{regular expressions (regexes)}, which enable precise matching of attack traces across heterogeneous log data. Regexes support a wide range of tasks, including digital forensics, log parsing, and the construction of rules in Security Information and Event Management (SIEM) systems. However, regex construction remains largely manual: analysts must extract IOCs from CTI reports and then craft syntactically valid and semantically precise patterns. This process is slow, error-prone, and increasingly unsustainable given the rising volume of CTI reports.
While recent approaches have applied Large Language Models (LLMs) to IOC extraction, they primarily focus on producing raw artifacts—leaving IOCs as plain strings that still require manual transformation into regexes before they can be operationalized. This gap limits their practical utility, as plain strings cannot account for variations in system contexts, log formats, or attacker behaviors, all of which regexes are designed to capture.
To address these challenges, we propose a fully automated LLM-based regex generation system, IOCRegex-gen, that generates corresponding regexes from IOCs. Our system introduces two key innovations: (i) a group-aware mechanism that automatically distinguishes which IOC segments should be treated as capture groups versus non-capture groups, and (ii) an iterative reasoning and multi-stage validation pipeline that ensures regexes are both syntactically valid and semantically correct.
Our experimental validation using over 3,000 real CTI reports and over 2,400 ground truth strings collected from the MITRE ATT\&CK Evaluation framework demonstrates that the regexes generated by our system achieve an average hit rate of 99.1\% across the datasets, with an average false‑positive rate of only 0.8\%, confirming its practical utility for large-scale CTI processing and automated regex generation.  
\end{abstract}

%
\IEEEpeerreviewmaketitle
\section{Introduction}
\label{introduction}
In modern Security Operations Centers (SOCs), analysts rely on a wide variety of data sources collected on a daily basis, including CTI reports, Windows event logs, endpoint logs, and Linux journals. To utilize these data sources in a scalable way, they must be parsed automatically to find the specific information items that analysts are searching for.
To serve the purpose of IOC hunting, CTI reports are a valuable resource, as they describe attacker behaviors and provide IOCs such as file paths, registry keys, and command-line arguments.
However, there is a fundamental gap between how attacker behaviors are described in CTI reports and how they are recorded in specific system logs. 
For example, a CTI report may describe an attacker creating a scheduled task with the command \texttt{schtasks.exe \textbackslash create \textbackslash tn …}, but in system logs, the same activity may appear with different path delimiters, execution parameters, or capitalization, making exact string matching infeasible. 
Without bridging this gap, even if IOCs are successfully extracted from CTI reports, threat intelligence remains difficult to operationalize in practice. 
The de facto strategy to bridge this gap is \textbf{the use of regexes}. In practice, human analysts manually translate IOCs extracted from CTI reports into regexes, which can then be executed by computers to search massive log data. Regexes thus provide the medium that operationalizes human-readable intelligence into actionable detection, supporting tasks such as digital forensics, log parsing, and the construction of SIEM rules.

When constructing regexes for IOCs, security analysts typically employ a \textbf{multi-step workflow}. They first review CTI reports from cybersecurity companies, individual experts, or knowledge bases such as MITRE ATT\&CK, extract the relevant IOCs, and then manually translate them into regexes. 
Although this manual process provides a \textbf{very important} enabler for many advanced SOC operations, it faces critical limitations that have grown more severe as the frequency and scale of global attacks increase. \cite{ctimarket}.
In practice, the current multi-step workflows face several critical limitations.
First, the end-to-end process of reading CTI reports, authoring regexes, and subsequently testing and tuning them can take hours to days, creating bottlenecks that delay responses to emerging threats. Second, regex authoring expertise is unevenly distributed: novice analysts often resort to senior colleagues for help, straining already limited security team resources. Third, the repetitive and detail-oriented nature of regex construction contributes to analyst fatigue, leading to errors such as syntax mistakes, missing capture groups, or insufficient generalization. These limitations introduce mounting time and labor pressures on SOC operations.

Recent research alleviates these limitations by automating part of the multi-step workflow. 
One line of work focuses on IOC extraction from CTI reports. For instance, \cite{Rastogi} utilized a pre-trained named-entity recognition (NER) model to extract IOCs from text and constructed a graph to represent inter-IOC relationships, while \cite{Husari2017} applied a pre-trained model to identify candidate threat actions for timely cyber defense. More recently, \cite{HU2024103999} and \cite{Schwartz2024} employed LLMs to extract IOCs from CTI texts. While these extraction approaches are valuable, their functionalities stop at producing raw IOCs and do not address the subsequent—and equally critical—step of translating these IOCs into deployable regexes. Consequently, analysts must still manually perform this regex construction, leaving the core operational bottleneck unresolved.

Beyond IOC extraction, several recent studies \cite{Xu2024, Schwartz2024, Wang2026RulePilot} have attempted to take one further step by leveraging LLMs to generate SIEM detection rules from CTI data automatically. However, in these works, the proportion of generated rules that incorporate regexes remains unrealistically small. For instance, \cite{Xu2024} utilized the in-context learning capabilities of LLMs to perform Tactics, Techniques, and Procedures (TTPs) classification and generate detection rules informed by external threat intelligence databases. While their results include 135 automatically generated rules, the vast majority correspond to surface-level indicators such as IP addresses and domain names, with only 7 rules incorporating any form of regex-based pattern matching, and these regex components are very simple in terms of structure and significantly simpler than the regexes in real-world SIEM rules, reflecting very limited exploration of regex synthesis in current approaches. 

Another line of work has explored regex generation in different contexts, though not specifically for security IOCs. These approaches can be broadly categorized into two groups. 
Traditional Natural Language Processing (NLP) methods, such as \cite{locascio2016neuralgenerationregularexpressions} and \cite{chen2020multimodalsynthesisregularexpressions}, improved syntactic validity through sequence-to-sequence models but were designed for controlled natural language tasks with well-formed input descriptions. More recently, LLM-based methods such as \cite{zhang2023inferestepbystepregexgeneration} and \cite{TangYLDCG24} have extended regex generation capabilities through improved reasoning and multi-step inference.  
However, these approaches remain centered on general natural language-to-regex translation and lack mechanisms to handle the unique challenges of security contexts: IOC variants exhibit adversarial variations intentionally designed to evade detection; there is a requirement for balancing precision with generalization to catch attack variations; and there is a requirement to maintain operationally meaningful capture groups for downstream SIEM integration.
Furthermore, unlike general regex generation tasks, where training examples are abundant, real-world IOCs are scarce, sensitive, and constantly evolving, making it infeasible to collect large sets of positive and negative examples for data-driven learning approaches such as supervised training or model fine-tuning. Consequently, a clear gap exists between the existing regex generation approaches and the requirements for dealing with the adversarial and resource-constrained nature of security operations.

In this work, we aim to take the critical remaining step of translating raw IOCs into deployable regexes. As implied by the above discussion, this step involves significant challenges. (\textit{C1}) Regex construction requires determining which substrings of an IOC should serve as capture groups for extracting key threat indicators. Misplacing capture groups not only weakens detection accuracy but also disrupts downstream integration with SIEM rules and forensic analysis tools, where extracted values are used for correlation and investigation.
(\textit{C2}) Regex generation with LLMs is prone to hallucinations and errors—models may produce syntactically invalid patterns or semantically misaligned expressions that fail to match the intended IOC behavior. Without robust mechanisms for validation and correction, such outputs cannot be safely deployed in production environments.
(\textit{C3}) Regex generation must balance precision and generality: overly narrow patterns miss true attack actions, while overly broad ones—such as those dominated by wildcards or optional groups—yield little practical detection value.

To address these challenges, we propose \textbf{IOCRegex-gen}, a novel LLM-based system that automates the translation from extracted IOCs to deployable regexes—the critical missing step in the multi-step workflow. Our system accepts IOCs extracted by any existing method as input and addresses the three challenges through two core components.
The first component employs a knowledge-enhanced framework integrated with a graph database to identify which substrings of each IOC should be designated as capture groups (addressing \textit{C1}). By retrieving relevant threat intelligence context and analyzing structural patterns across similar IOCs, this component ensures that capture groups align with operational requirements for SIEM integration and forensic analysis.
The second component implements a reasoning-based generation process with iterative validation to ensure both syntactic correctness and semantic accuracy (addressing \textit{C2} and \textit{C3}). Through multi-stage validation checks and refinement loops, the system mitigates LLM hallucinations, prevents the generation of syntactically invalid patterns, and maintains an appropriate balance between precision and generality by rejecting overly broad or overly narrow regex candidates.

To validate our system's effectiveness, we conduct experiments using the MITRE ATT\&CK Evaluation framework \cite{mie}, which provides independently collected artifacts from actual security incidents. Specifically, the independently collected artifacts are reported by cybersecurity companies during these MITRE ATT\&CK evaluation exercises. We let these artifacts serve as \textbf{ground-truth}.
We then generated regexes from over 3,000 real-world CTI reports using our system and evaluated their performance against these ground-truth strings.
Our experimental results demonstrate that IOCRegex-gen achieves strong detection performance: the generated regexes attain an average hit rate of 99.1\% on ground-truth strings while maintaining a low false-positive rate of 0.8\%.  
Furthermore, quantitative analysis shows that the generated regexes achieve an average grading score above 3, indicating substantial structural complexity beyond simple literal matches. The average similarity score between generated regexes and their source IOCs is around 0.4, suggesting that the patterns retain essential IOC features while introducing sufficient variability for broader applicability.
Ablation experiments further confirm the importance of each system component. Compared to directly prompting an LLM to generate regexes without our specialized components, IOCRegex-gen achieves over 30\% improvement in both hit rate and false-positive rate, demonstrating that our design effectively addresses the challenges inherent in CTI-to-regex translation.

In summary, we have made the following contributions:
\begin{itemize}[nosep,leftmargin=2.2ex]
    \item We propose \textbf{IOCRegex-gen}, an LLM-based system that automates the translation from extracted IOCs to deployable regexes, addressing the critical missing step in the CTI operationalization workflow and bridging the gap between threat intelligence extraction and actionable security detection.
    \item We develop novel methods for automated capture group identification through knowledge-enhanced context retrieval and graph-based pattern analysis, coupled with a reasoning-driven generation workflow that performs iterative validation to ensure syntactic correctness, semantic accuracy, and appropriate generalization.
    \item We conduct a comprehensive evaluation using over 3,000 real-world CTI reports and ground-truth strings from the MITRE ATT\&CK Evaluation, demonstrating that our system achieves a 99.1\% hit rate with a 0.8\% false-positive rate and outperforms direct LLM baselines by over 30\% in both metrics.
\end{itemize}

\noindent\textbf{Open Source}: The source code for our implementation is available at \url{https://anonymous.4open.science/r/IOCRegex-gen-8FFE/README.md}  
\section{Background}
\subsection{Cyber Threat Intelligence}
\label{Background1}
CTI reports serve as an essential medium for cybersecurity analysts to exchange threat intelligence, providing detailed insights into attackers’ methods, workflows, and technical procedures. For example, \cite{TeamTNT} documents the background, targets, and tactics of the hacker group TeamTNT. While some CTI reports are paid resources, many are freely accessible and contributed by both organizations and individuals. Major cybersecurity vendors such as Trend Micro and FireEye routinely publish high-quality CTI reports \cite{RedCurl, TeamTNT, TRITON}, whereas individual researchers and SOC analysts often share technical analyses on blogs or personal platforms \cite{RedCurl2, Astaroth}. In addition, centralized repositories like MITRE ATT\&CK \cite{123} and vendor-specific platforms such as Trend Micro’s Threat Encyclopedia \cite{threat-encyclopedia} aggregate CTI content for broader accessibility. 
\vspace*{-2mm}
\subsubsection{Regex-Applicable IOCs}
\label{high_value_ioc}
The IOCs embedded within CTI reports serve as clear indicators to determine whether a system has been breached. However, not all types of IOCs are suitable for regex generation. According to the Pyramid of Pain \cite{Pyramid}, the low-value indicators—such as hash values, IP addresses, domain names, and network artifacts—are the ones most easily changed by adversaries. These indicators are highly volatile, as attackers can regenerate file hashes, rotate or proxy IP addresses, and frequently modify domain names or network configurations. They are also instance-specific and lack consistent syntactic structure, which makes it difficult to generalize them into meaningful regexes. For example, a hash value such as \texttt{d41d8cd98f00b204e9800998ecf8427e} is unique to a single file and provides no reusable pattern that a regex could capture. In practice, these indicators are more effectively handled through exact-match lookups, threat intelligence feeds, or reputation-based systems rather than pattern matching. 

\begin{figure}[!t]
\centering
\includegraphics[width=0.7\linewidth]{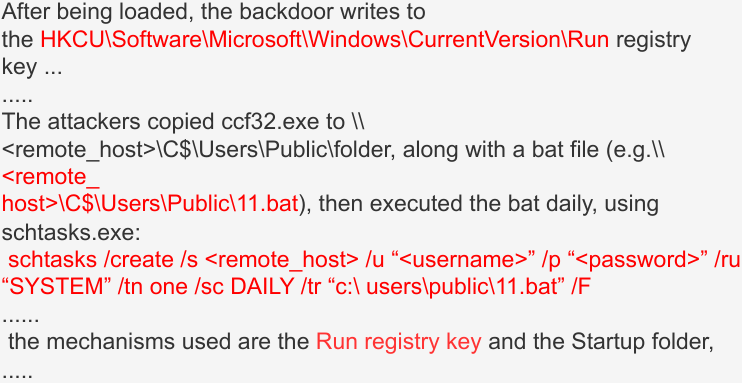}
\caption{Example paragraphs from a CTI report}
\label{fig:example_of_cti}
\vspace{-4mm}
\end{figure}

In contrast, \textbf{file paths, registry keys, and command-line arguments} represent the most critical and \textbf{regex-applicable types of IOCs}, since they reliably reveal persistent threats and inherently contain both fixed and variable components—such as stable system directories combined with attacker-controlled filenames or parameters. These structural characteristics make them difficult to operationalize through exact matching but highly suitable for regex-based representation, which is essential for enabling their use in large-scale searching and SOC tasks. In Figure \ref{fig:example_of_cti}, the text illustrates how the attacker leverages the Chinoxy backdoor to achieve persistence on the victim machine and to schedule regular execution of a data-collection utility. First, the attacker drops the backdoor binary into the victim’s user-profile directory and writes its full path to the registry key \texttt{HKCU\textbackslash Software\textbackslash Microsoft\textbackslash Windows\textbackslash CurrentVe\\rsion\textbackslash Run}. Thereafter, whenever the user logs into Windows, the backdoor automatically launches, ensuring it remains resident. In addition, the attacker uses Windows’ built-in scheduling tool \texttt{schtasks.exe} to create a daily task on the remote host that executes the data-collection program. In this text, the registry key entries, the path to the malicious executable, and the attacker’s command line argument invocations constitute critical indicators of attack. When these indicators appear on a system, they may signify that the system is under attack or has already been compromised. Security analysts closely scrutinize CTI reports to identify IOCs, as these indicators are essential for establishing SIEM rules or guiding digital forensics.
\vspace{-2mm}
\subsubsection{IOC Extraction}
\label{ioc_extraction}
Extracting IOCs from CTI reports is an essential step in transforming unstructured textual threat descriptions into structured and actionable data. Traditionally, analysts manually review CTI reports to locate IOCs. However, as the scale and complexity of CTI reports continue to grow, manual extraction has become increasingly time-consuming and prone to inconsistency, motivating the development of automated approaches.

Recent studies have explored diverse NLP-based approaches to automate IOC extraction. Some works employ pre-trained models for named entity and relation extraction to identify threat entities and contextual relationships in CTI narratives \cite{Husari2017, Rastogi, Gao}. Others LLMs to directly interpret CTI text, extracting not only IOCs but also associated TTPs, and constructing structured threat knowledge graphs \cite{hu2024llm, huang2024ctikg, liu2023constructing, cuong2025towards, fieblinger2024actionable}. For instance, \cite{hu2024llm} annotates entities and relationships to enhance downstream model training, and \cite{huang2024ctikg} builds CTI-oriented knowledge graphs to represent threat behaviors and mitigate hallucinations.

Despite such progress, existing work treats IOC extraction as an intermediate to serve non-regex-construction purposes, such as knowledge graph construction or TTPs identification. While these approaches greatly advance the structuring and interpretation of CTI, the operational use of extracted IOCs—for example, transforming them into deployable regex for digital forensics—remains less explored in the current literature.

\vspace{-7mm}
\subsection{Regular Expressions in Security Operations}
\vspace{-7mm}
\label{The role of regular expression in SOC's security analysis}

In SOC operations, regex is commonly used because, despite the usefulness of IOCs from CTI reports, they typically cannot be directly matched with system logs for two key reasons. First, each machine’s system context is unique. IOCs may reflect the specific context in which the CTI report was written, so even if the attacker uses the same command line argument, the actual implementation may vary slightly across different environments. For example, when an attacker uses the command line argument in Figure \ref{fig:example_of_cti} to create a scheduled task, they might change the path of the invoked program depending on the machine's system context, or adjust the execution time based on the machine’s timezone. Furthermore, relying on exact matches against raw IOCs is impractical; even slight differences in capitalization, path delimiters, or whitespace will often cause missed detections. If pattern matching relied solely on literal strings, variations like \texttt{temp1.exe}, \texttt{temp2.exe}, \texttt{tmpA34.dll} would evade detection due to their minor naming differences. Using regex, however, allows these payloads to be matched collectively by identifying shared elements like filename prefixes and extensions. Therefore, after obtaining IOCs, security analysts construct regexes to efficiently identify string variations that are similar to, but not exactly the same as, the original IOCs—making large-scale IOC searching in massive log datasets both feasible and reliable. In practice, the widespread adoption of regex across commercial SIEM platforms further underscores its operational importance in SOC workflows. 

Many SIEM systems natively employ regex, enabling analysts to flexibly filter complex log patterns and reduce irrelevant results. Splunk provides ``rex'' and ``regex'' commands for direct pattern searches \cite{splunk_regex}, Elastic SIEM supports regex in its query language for large-scale indexed data \cite{elastic_regex}, and IBM QRadar integrates regex into its property extraction and correlation rules \cite{qradar_regex}. Moreover, we analyzed 1,735 public Splunk detection rules \cite{Splunk_Security_Content} and found that over 64\% contained regex patterns related to file paths, registry keys, or command-line arguments, indicating that these elements frequently serve as matching targets in practical SOC detection logic.

\vspace*{-2mm}\section{Motivation}
\label{motivation}
\subsection{SOC Regular Expression Generation Workflow}
\vspace*{-2mm}
\label{motivation1}
In real-world SOCs, security analysts typically manually generate regexes. In particular, the workflow is as follows: upon receipt of a CTI report (which can range from a few pages to several thousand words), the analyst first reviews the document to determine which techniques the adversary employs, and then extracts a raw list of IOCs from the narrative descriptions of those techniques. Next, the analyst manually deduplicates overlapping entries, filters out low‑value IOCs, and retains the high‑value IOCs mentioned in Section \ref{high_value_ioc}.

Building on the remaining IOC list, the next step for a SOC analyst is to decompose each IOC into its mutable and immutable elements, denoted in regex as non-capture groups and capture groups, respectively. \textbf{Non-capture groups} represent segments of an IOC that can change between machine's system context or due to attacker variation, such as randomized file names, GUIDs, or ephemeral registry key values, while \textbf{capture groups} correspond to constant, unchanging components like system-built directory paths or built-in command names, which threat actors habitually use to make their behavior appear normal\cite{Fileless}. For example, given the file path IOC \texttt{C:\textbackslash Users\textbackslash Public\textbackslash 11.bat}, an analyst might define the path prefix \texttt{Users\textbackslash Public} as a capture group, and the payload name \texttt{11.bat} as a non-capturing group. Security analysts identify these parts by combining their institutional knowledge of OS conventions and common adversary tradecraft with external documentation, such as Microsoft’s Sysinternals guides or CTI vendor IOC repositories, to determine which elements truly require flexibility. Once grouping is delineated, they construct an initial regex that combines literal matches, quantifiers, and alternations (e.g. \texttt{.*\textbackslash Users\textbackslash Public\textbackslash.*}), then iterate on it using regex testing tools like regex101 \cite{regex101} to confirm it’s syntactically valid, not overly permissive, and precise enough to catch IOC variants without false positives. Only once the regex pattern passes the validation is it deployed into the SOC’s SIEM system.

\vspace{-2mm}
\subsection{LLM-based system to Automate the Workflow}
\vspace{-2mm}
\label{motivation2}
\begin{figure}[t]
\centering
\includegraphics[width=0.5\textwidth]{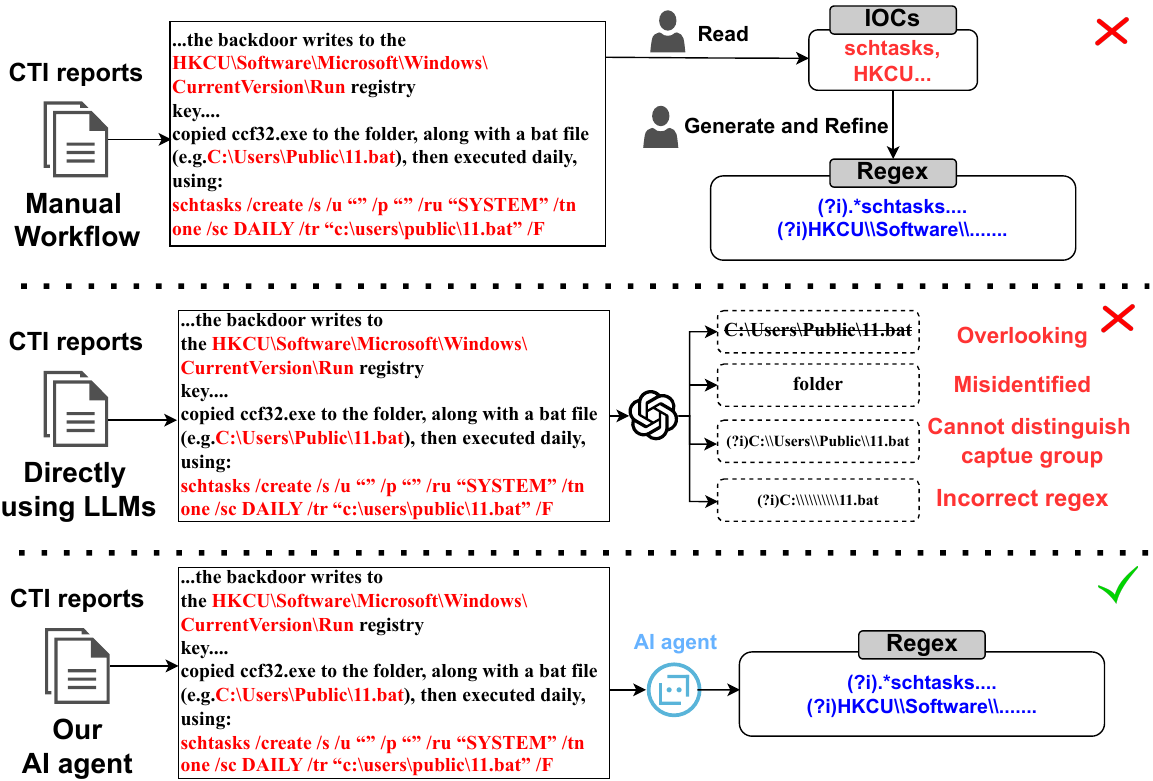}
\caption{Motivating Example}
\label{fig:fig4}
\vspace{-4mm}
\end{figure}
By leveraging an LLM‑based system, the regex generation workflow could be automated through the inherent capabilities provided by LLMs. However, utilizing LLMs to complete this task presents several challenges that must be overcome. As discussed in Section~\ref{ioc_extraction}, prior research has primarily focused on automatic IOC extraction. In this section, we focus on the remaining challenges specific to the regex generation workflow.

First, current LLMs lack sufficient knowledge to distinguish which parts of an IOC belong to the capture group versus the non-capture group (\textit{C1}). Unlike experienced analysts, who rely on domain knowledge of operating system conventions, LLMs often struggle with this distinction. For instance, using the example mentioned in Section \ref{motivation1}, where the IOC \texttt{C:\textbackslash Users\textbackslash Public\textbackslash 11.bat} could also appear as \texttt{12.bat}, a human analyst would recognize that the stable path prefix \texttt{Users\textbackslash Public} should be treated as a capture group, while the filename is attacker-controlled and better represented as a non-capture group. LLMs, by contrast, lack the contextual understanding to make this separation reliably. To address this limitation, our system employs a knowledge-enhanced context retrieval method, which will be shortly presented in Section~\ref{Capture_Group_Finding}. This method enriches the capture group finding process with structured system knowledge that reflects the intrinsic organization of operating environments. Specifically, it incorporates canonical information about file path hierarchies, registry key structures, and command-line utility patterns—elements that remain consistent across machines and provide stable references for determining which IOC components should be treated as capture groups. 

Second, LLMs frequently fail to produce syntactically valid and complete regexes, particularly when multiple capture groups or nested patterns are involved (\textit{C2}). Although LLMs have demonstrated strong capabilities in logical tasks such as mathematical reasoning and code generation, their application to regex synthesis remains largely unexplored. Our observations show that LLMs frequently produce invalid or incomplete regexes when tasked with longer or more intricate patterns, and they often omit required capture groups specified in the prompt. Previous studies have primarily applied reasoning-based methods to domains such as mathematical problem solving and code generation, where model feedback and external tools are used to iteratively refine outputs toward the desired solution \cite{yao2023reactsynergizingreasoningacting,wei2023chainofthoughtpromptingelicitsreasoning}. In our system, we implement a dedicated reasoning workflow for regex generation: using this method, the LLM incrementally constructs a correct and valid regex step by step, ensuring that the required capture groups are preserved.

Third, even when LLMs generate syntactically valid regexes, these patterns can often be overly general and thus useless in practice (\textit{C3}). For instance, the model may produce regexes dominated by wildcards or optional groups, such as \texttt{(...)?}, resulting in patterns that technically compile but match almost everything, thereby providing no discriminative power. To mitigate this issue, our system builds upon the reasoning-based method described above to guide the model away from excessively permissive constructions. In addition, we introduce a grading system that evaluates candidate regexes against a set of criteria and discards those that fail to provide meaningful specificity, thereby ensuring that only regexes with practical detection value are retained.

\section{Approach}
\begin{figure*}[!h] 
\centering
\includegraphics[width=0.9\textwidth]{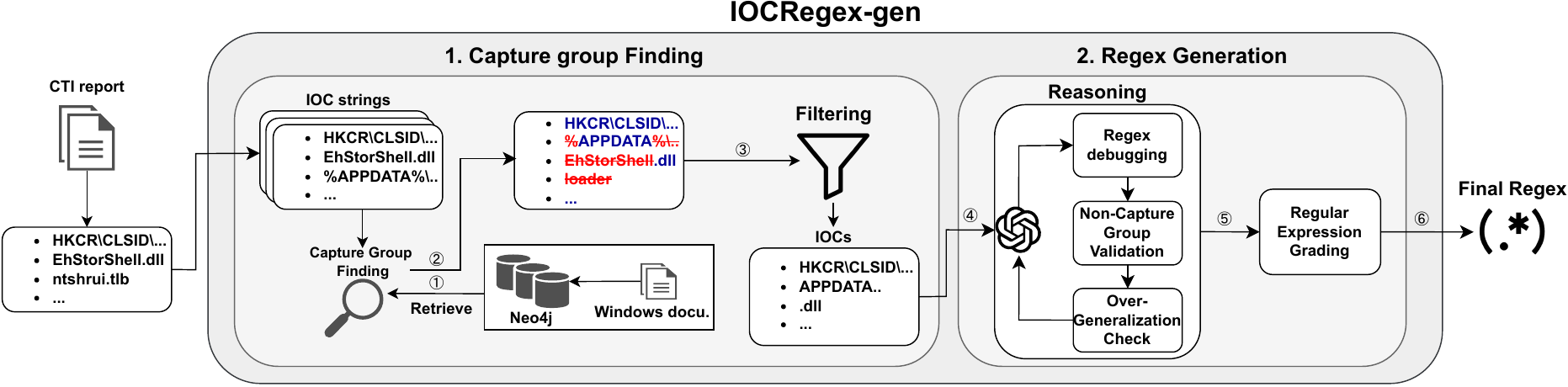}
\caption{Overview of IOCRegex-gen}
\label{fig:fig2}  
\end{figure*}
We propose IOCRegex-gen, a system that transforms raw IOCs from CTI reports into syntactically valid and semantically precise regexes. Since our goal is to automate the translation of IOCs into regexes, the system focuses exclusively on \textbf{regex-applicable IOC} mentioned in Section \ref{high_value_ioc}, and accepts IOC outputs from any existing extraction framework. Figure \ref{fig:fig2} provides an overview of the overall system. Generally, IOCRegex-gen can be divided into two phases. In the Capture Group Finding phase, which addresses the first challenge (\textit{C1}) discussed in Section \ref{motivation2}, \ding{172} IOCRegex-gen retrieves from a pre-established graph database the relationships between substrings in the potential IOC strings, \ding{173} applies algorithms to determine which substrings are capture groups, and then, using a designed filtering mechanism, \ding{174} removes false positives identified in the IOC extraction. The subsequent Regex Generation phase targets the second challenge (\textit{C2}) and third challenge (\textit{C3}) identified in Section \ref{motivation2}. During this phase, \ding{175} IOCRegex-gen uses a tool to evaluate the LLM’s proposed regex against the target IOC. The tool’s output, which details any discrepancies between the suggested regex and the IOC, is then used to craft the next prompt to guide the LLM toward an improved regex. This iterative reasoning process continues until the tool reports no discrepancies. At the same time, the reasoning process also ensures that the regex is not over-generalized and does not include any non-capturing groups. Finally, \ding{176} IOCRegex-gen leverages the above‐mentioned reasoning process to repeatedly generate multiple regexes for the same IOC. \ding{177} A scoring mechanism is then applied to evaluate these candidates and select the highest‐scoring regex. The selected pattern becomes the final regex for that IOC.

\vspace*{-2mm}
\subsection{Capture Group Finding}
\vspace*{-2mm}
\label{Capture_Group_Finding}
Building on the definitions introduced in Section \ref{motivation}, our regex generation pipeline must include a dedicated grouping step that isolates invariant IOC fragments from those likely to vary across environments or attacker campaigns. 

In this phase, IOCRegex-gen retrieves data from an external graph database and uses our proprietary algorithm to differentiate between capture groups and non-capture groups within each string extracted during IOCs Extraction. Additionally, when encountering any string that does not contain a capture group, it is considered a false positive from the previous stage and is subsequently removed, and only the strings that remain after filtering are considered genuine IOCs.

\subsubsection{\textbf{Graph Database}} 
\begin{figure}[!t] 
\centering
\includegraphics[width=0.7\linewidth]{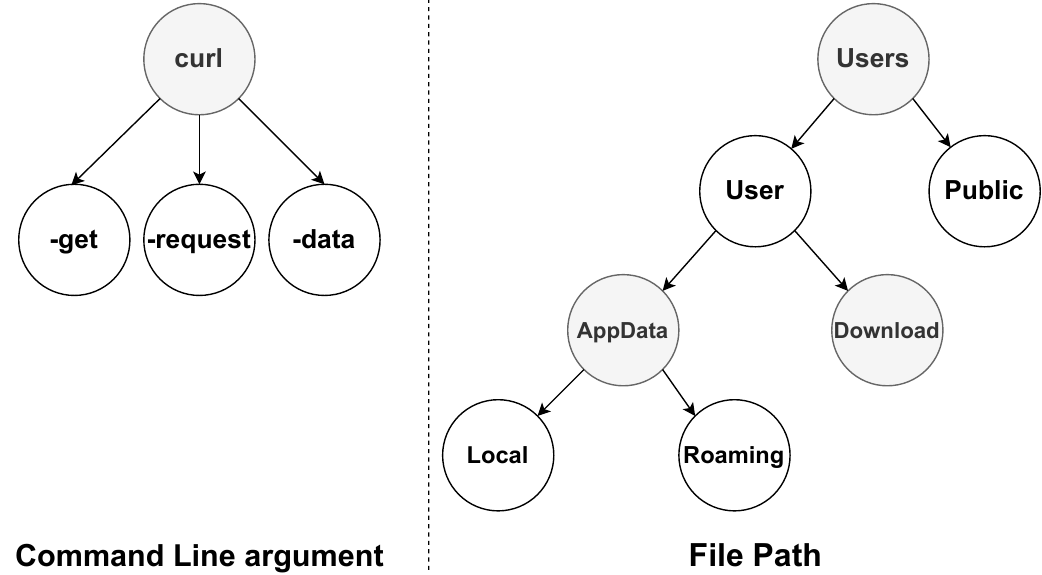}
\caption{The tree structure of command line argument and file path}
\label{fig:database} 
\vspace{-4mm}
\end{figure}

As mentioned in Section \ref{motivation1}, most attackers tend to use the operating system’s built-in command line arguments, registry keys, or directories to conceal their malicious activities. These built-in components are treated as capture groups. Therefore, we have collected all native file paths, registry keys, and command line arguments for Windows 8, 10, and 11, as well as Windows Server 2012, 2016, 2019, and 2022. To facilitate efficient retrieval of data for capture group finding in section \ref{matching}, this data is stored in the graph database as a tree structure. We create the tree structure for file paths, registry keys, and command line arguments separately using the following methods. 1) We split file paths or registry keys using slashes as delimiters to obtain substrings. In the tree structure, each substring is treated as an individual entity, and each entity is connected by a link established according to the hierarchical order, effectively associating them with one another. For example, as shown in Figure \ref{fig:database}, under the \texttt{Users} root directory, there may be several subdirectories named after individual users. Here, the \texttt{Users} directory serves as the parent node, its subdirectories are child nodes, and each of those child nodes may in turn contain further subdirectories as their own child nodes. 2) When constructing the tree structure for the command line arguments, we first determine which element represents the command and which ones are the subordinate parameters. For example, \texttt{curl} is a tool used for sending and receiving various network requests. It has many parameters, such as \texttt{--get}, \texttt{--request}, \texttt{--data}, and others. We then establish parent and child relationships from curl to its parameters in the tree structure, reflecting their hierarchical dependency.

\subsubsection{\textbf{Preprocessing}} Typically, CTI reports are written by humans without a standardized format, so it is necessary to standardize the extracted strings. For instance, a registry key’s root (e.g., \texttt{HKEY\_CURRENT\_USER}) might appear as \texttt{HKCU} or simply as \texttt{REGISTRY}. Similarly, when representing a user's personal folder, the path might be expressed as \texttt{C:\textbackslash Users\textbackslash \textless username\textgreater}, \texttt{\textless username\textgreater}, or even \texttt{user}.

During preprocessing, We first classify the string obtained from IOCs Extraction based on its structure to determine whether it is a file path, registry key, or command line argument, and then handle several common cases:  1) Environment Variable Expansion: For file paths that may be expressed using environment variables, such as \texttt{\%USERPROFILE\%}, we restore them to their original form. 2) Username Normalization: Various representations of usernames are standardized by replacing them with a uniform token (e.g., \texttt{user}). 3) Registry Key Standardization: For all registry keys, the root keys are standardized by using their abbreviated forms. 4) Command Line Argument Normalization: For command line arguments, the commands (e.g., \texttt{cmd.exe}, \texttt{curl.exe}, or \texttt{wmic.exe}) are normalized by removing file extensions.

\vspace*{-2mm}
\subsubsection{\textbf{Capture Group Finding}}\label{matching}After preprocessing the potential IOC strings, file paths, and registry keys are split into individual substrings using slashes, while command line arguments are split based on whitespace, semicolons, and other delimiters to extract substrings. Since the structure of command line arguments differs from that of file paths and registry keys, we employ two distinct algorithms to identify capture groups within these strings. 

In Algorithm \ref{algo1}, the input $S$ is a list of substrings that are ordered according to their hierarchical position within the path, and $D$ represents the graph database. Additionally, $V$ is defined as the set of substrings in list $S$ that appear as entities in the graph database. Let $t$ be an entity in the graph database; if $t$ is adjacent to $s$ in the graph, we denote this relationship as $sAt$. For a given list $S$, we start by verifying which substrings in the list appear in the graph database (line 1). This initial step filters out substrings that are clearly not native to the operating system, for example, file or folder names arbitrarily assigned by an attacker. Next, we take the first remaining substring and add it to the current sequence $C$. We then check if the next substring is adjacent to the previous one in the $D$. If they are adjacent, we add the substring to $C$ and continue checking the following substrings in the same manner (lines 3-8). If a substring is not adjacent to the previous one, we record the length of the current sequence C and restart the process with the non-adjacent substring as the new starting point. This iterative process continues until we identify the longest sequence, L, of adjacent substrings (lines 6-11). We retain only the longest sequence as the capture group because we've observed that some directories native to the operating system often recur across different file paths. Additionally, this approach helps prevent capturing directory names that may have been arbitrarily defined by attackers. 
    \begin{algorithm}[htbp]
    \footnotesize
    \caption{Capture group finding for file path and registry key}
    \label{algo1}
    \KwIn{$S = \langle s_1, s_2, \dots, s_n \rangle,\quad s_i \in \Sigma$}
    \KwIn{$D$, the graph database}
    \KwData{$V \gets \langle s \in S \mid s \in D \rangle$}
    \KwData{$s\,A\,t \iff t \text{ is the adjacent node of } s \text{ in } D$}
    \KwOut{$L = \langle \ell_1, \ell_2, \dots, \ell_k \rangle,\quad k = \max\{\,|C| : C \subseteq V,\; \forall\, j,\; V_j\,A\,V_{j\texttt{+}1}\,\}$}
    \BlankLine
    $V \gets \langle s \in S \mid s \in D \rangle$\;
    $max \gets 0$, $L \gets \langle\,\rangle$\;
    \For{$i \gets 1$ \KwTo $|V|$}{
        $C \gets \langle V_i \rangle$\;
        $j \gets i$\;
        \While{$j < |V|$ \textbf{ and } $V_j\,A\,V_{j\texttt{+}1}$}{
            $C \gets C \circ V_{j\texttt{+}1}$\;
            $j \gets j\texttt{+}1$\;
        }
        \If{$|C| > max$}{
            $max \gets |C|$, $L \gets C$\;
        }
    }
    \Return $L$\;
    \end{algorithm}
For example, consider the file path \texttt{C:\textbackslash Users\textbackslash Public\textbackslash 11.bat} in Figure \ref{fig:example_of_cti}. The algorithm first checks whether the directory \texttt{Users} exists as a node in the graph database. If it does, it then verifies that \texttt{Public} is a child node of \texttt{Users} in the tree structure. Next, it checks whether \texttt{11} is a child node under \texttt{Public}. If it is not, then the capture group for that file path is \texttt{Users\textbackslash Public}.

In Algorithm \ref{algo2}, the input consists of a list $S$ of substrings along with the graph database $D$. However, the output is a set of sequences $\Omega$ because a single command line argument may contain multiple main commands. Given a list $S$ containing multiple ordered substrings, the algorithm starts with the first substring and verifies whether it exists in $D$ and belongs to the command category. If so, it is added to the current sequence $C$ (lines 5-10). The algorithm then checks whether the subsequent substring exists in $D$ and belongs to the parameter category. If it does and is adjacent to the previous substring in the graph, it is added to $C$; otherwise, it is discarded (lines 11-12). This process is repeated until the next substring identified as a command is encountered (lines 5-14). The final output is a collection of sequences $\Omega$, where these sequences represent the capture groups for the string related to the command line argument.
\begin{algorithm}[htbp]
\footnotesize
\caption{Capture group finding for command line argument}
\label{algo2}
\KwIn{$S = \langle s_1, s_2, \dots, s_n \rangle,\quad s_i \in \Sigma$}
\KwIn{$D$, the graph database}
\KwData{
\(G(s,\mathrm{label})\) returns true if \(s\) is marked as the given label (\(\mathrm{command}\) or \(\mathrm{parameter}\)) in \(D\). \\}
\KwData{$s\,A\,t \iff t \text{ is the adjacent node of } s \text{ in } D$}
\KwOut{\(\Omega = \langle \omega_1, \omega_2, \dots, \omega_k \rangle\), where each \(\omega\) is an ordered sequence of substrings starting with a command followed by its parameters}
\BlankLine
$V \gets \langle s \in S \mid s \in D \rangle$\;
\(\Omega \gets \langle\,\rangle\)\;
\(C \gets \langle\,\rangle\) \\
\(c \gets \text{undefined}\) \\
\For{\(i \gets 1\) \KwTo \(|V|\)}{
    \If{\(G(V_i,\text{``command''})\) is true}{
        \If{\(C \neq \langle\,\rangle\)}{
            \(\Omega \gets \Omega \circ \langle C \rangle\)\;
        }
        \(C \gets \langle V_i \rangle\)\;
        \(c \gets V_i\)\;
    }
    \ElseIf{\(G(s_i,\text{``parameter''})\) is true \textbf{and} \(c\,A\,V_i\)}{
        \(C \gets C \circ \langle V_i \rangle\)\;
    }
}
\If{\(C \neq \langle\,\rangle\)}{
    \(\Omega \gets \Omega \circ \langle C \rangle\)\;
}
\Return \(\Omega\)\;
\end{algorithm}
For instance, using the command line arguments \texttt{schtasks /create /s \textless remote\_host\textgreater /u "\textless username\textgreater" /p "\textless password\textgreater" /ru "SYSTEM" /tn one /sc DAILY /tr "c:\textbackslash users\textbackslash public\textbackslash 11.bat" /F} shown in Figure \ref{fig:example_of_cti} as an example, the algorithm first identifies the command, \texttt{schtasks}. It then uses the tree structure in the graph database to determine which substrings are child nodes of that command (e.g., \texttt{/create}, \texttt{/s}, \texttt{/u}, etc.), and simultaneously flags any substrings not found in the tree, such as \texttt{\textless remote\_host\textgreater}, \texttt{\textless username\textgreater}. Finally, it treats those substrings absent from the tree as non-capture groups, and all remaining substrings as capture groups.






\vspace*{-2mm}
\subsubsection{\textbf{False Positive Filtering}}\label{False positive filtering} After the capture group finding phase, IOCRegex-gen processes each string from IOCs Extraction by labeling every component as either \textit{keep} or \textit{discard}. Components marked as \textit{keep} are considered to be part of the string's capture group, while those labeled as \textit{discard} are treated as part of the non-capture group. Finally, if a string does not contain any capture group components, IOCRegex-gen treats it as a \textbf{false positive} from the IOCs Extraction phase and discards it.

\vspace*{-2mm}
\subsection{Regex Generation}
\vspace*{-2mm}
Prior studies have shown that regex generation can be automated through rule-based or NLP-based methods. However, these approaches are ill-suited for IOC-related regex generation. Methods such as \cite{osti_10336574} require large numbers of representative positive examples, which are difficult to obtain for IOCs. Rule-based approaches also struggle with the diverse formats and structures of IOCs. Moreover, since each IOC may involve different capture groups, existing methods lack the flexibility to generate accurate regexes across varying IOC types.

\subsubsection{\textbf{Reasoning-based Regex Generation}} \label{Reasoning-based regex generation} With the advent of LLMs, it initially appeared that these models could address these challenges by generating regex patterns for IOCs. Compared to using a smaller deep learning model to generate regexes \cite{9401951}, the massive parameter scales of LLMs provide significant improvements in textual and logical processing. However, we observed that regardless of the LLM (e.g., GPT-4o, DeepSeek-V3, or LLaMA3) used, none can consistently generate correct regex patterns for IOCs. In most cases, the regex output still contains non-capture groups, resulting in non-compliant regex patterns. Furthermore, we observed that even newer reasoning models, such as GPT-O4 or Deepseek-R1, which seem promising in solving these issues, struggle when faced with lengthy IOCs like complex command line arguments. In such cases, these reasoning models prove ineffective. To address these limitations, we propose a novel workflow that incrementally guides the LLM to generate compliant and accurate regex patterns. Remarkably, this workflow does not require the use of advanced reasoning models; using a standard LLM (e.g., GPT-4o) yields results that surpass those of current reasoning models. As shown in Figure~\ref{fig:fig3}, the process works as follows:

\begin{figure}[!htbp]
\centering
\includegraphics[width=0.6\linewidth]{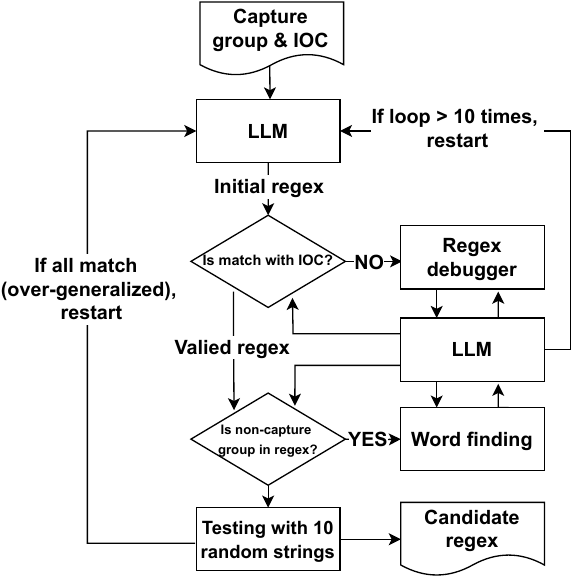}
\caption{Workflow of Reasoning-based Regex Generation}
\label{fig:fig3}
\vspace{-3ex}
\end{figure}

\vspace{1ex}
\noindent\textbf{Initial Regex Generation:} The IOC, along with the capture group information obtained from the previous phase, is fed into an LLM to generate an initial regex.

\vspace{1ex}
\noindent\textbf{Debugging.} The generated initial regex is tested to see if it can successfully match the original IOC. If it fails to match, a regex debugger is used to identify the exact step where the failure occurs. The debugger provides details on which parts of the regex can match and which cannot. This information is then used in a prompt to the LLM, which iteratively refines the regex until a valid one is obtained.

\vspace{1ex}
\noindent\textbf{Non-Capture Group Validation.} Once the regex has passed the debugging checks to confirm it still matches the original IOC, we run a second loop to ensure that it contains no capture groups. Within this loop, if any non-capture group components are found, relevant prompts indicating which non-capture groups are still present in the regex are fed back into the LLM along with the current regex, triggering further iterative refinement until a valid regex with no non-capture group elements is produced. If either the debugging loop or the non-capture group refinement loop exceeds ten iterations, it returns to the very beginning of the workflow to generate a new initial regex.

\vspace{1ex}
\noindent\textbf{Over-Generalization Check.} Even after obtaining a valid regex that matches the original IOC, it is sometimes observed that the LLM generates a regex composed entirely of wildcards(e.g., `.*' or `+w'), leading to over-generalization. To address this, in addition to the two loops mentioned above, we set up a third loop. In this loop, ten strings are randomly generated and tested to see whether the candidate regex matches all of them. If the regex matches every generated string, it is considered overly generic; it then restarts the workflow from the very beginning to create a new initial regex.

\vspace{-1mm}
\subsection{\textbf{Regular Expression Grading}} \label{Regular expression Grading} 
\vspace{-1mm}
In addition to the reasoning-based regex generation workflow described in Section  \ref{Reasoning-based regex generation}, we also employ a separate mechanism to ensure that the generated regexes genuinely include the intended capture groups and do not match any strings belonging to non-capture groups. 

Since the same IOC can be matched by multiple different regexes. Based on our observations, LLMs (e.g., GPT-4o) generate, on average, about five distinct regex variations. Exploiting this characteristic, we use the workflow from Section \ref{Reasoning-based regex generation} to generate five candidate regexes for each IOC, then score each regex using the formula below and retain only the highest-scoring regex as the final regex. In this scoring formula, 1) A regex receives points for each component that belongs to a capture group. 2) It incurs a penalty for each non-capture group included. The formula for scoring is as follows:
\begin{equation}
\label{eq:regex_generalization}
Score = \alpha\, n_{cg} - \beta\, n_{wc}
\end{equation}
$n_{cg}$ represents the number of components that belong to capture groups, and $n_{wc}$ represents the number of non-capture groups. Furthermore, the LLM may insert substrings into the regex that do not belong to any capture group; such substrings are also included in $n_{wc}$. The reward coefficient $\alpha$ and penalty coefficient $\beta$ determine the relative importance of capturing essential components versus avoiding substrings that should not be captured. We enforce a more stringent policy regarding the inclusion of non-capture groups. Therefore, we set $\alpha = 1$ and $\beta = 1$ to balance the scoring. It’s worth noting that if a capture group appears inside an optional group, such as `(...)?', we do not include it in the scoring.

\section{Evaluation}
\label{Evaluation}

\begin{table}[!t]
\centering
\label{tab:ioc_distribution}
\setlength{\tabcolsep}{3pt} 
\begin{tabular}{lcccc}
\toprule
\textbf{Type} & \textbf{File Path} & \textbf{Registry} & \textbf{CLI Args} & \textbf{Other} \\
\midrule
\textbf{Count} & 12{,}195 & 2{,}302 & 10{,}286 & 38{,}820 \\
\bottomrule
\end{tabular}
\caption{Distribution of extracted IOC types}
\vspace{-4mm}
\label{ioc_extrated}
\end{table}
Our evaluation addresses five research questions. \textbf{(RQ1)} Can IOCRegex-gen generate regexes in a scalable way? \textbf{(RQ2)} What are the false positive rates of the generated regexes? \textbf{(RQ3)} Are the generated regexes too general, causing unintended matches? \textbf{(RQ4)} Are the regexes general enough to detect attack pattern variations? \textbf{(RQ5)} What is the individual contribution of each system component?
\vspace*{-2mm}
\subsection{Experimental Setup} 
\vspace*{-2mm}
In this experiment, we used GPT-4o, LLaMA3, and DeepSeek-V3 to extract the IOCs and employed GPT-4o to generate the regexes, while the graph database was implemented using Neo4j. Additionally, all datasets and intermediate process data were stored in JSON format. The entire system is built with over 4000 lines of Python code.
\vspace*{-2mm}
\subsubsection{IOC Dataset for Regex Generation} \label{dataset} We collected \textit{3,156 CTI reports} from the references listed under the MITRE ATT\&CK framework \cite{123}. Each ATT\&CK technique web-page contains multiple external references, which we systematically crawled and parsed to obtain a diverse corpus of CTI reports.

To ensure a consistent and comprehensive dataset of IOC inputs for evaluation, we adopted an extraction process inspired by recent LLM-based CTI parsing research\cite{Rastogi, wang2023selfconsistency, fieblinger2024actionablecyberthreatintelligence}.
The process combines three complementary techniques to maximize recall and ensure practical applicability in large-scale CTI analysis.
First, a semantic-based chunking method is applied to divide long CTI reports into coherent and manageable segments, similar to \cite{Rastogi}, which demonstrated that segmentation preserves contextual integrity and prevents extraction drift.
Second, multiple LLMs—including GPT-4o, DeepSeek-V3, and LLaMA3—independently analyze each segment. Following the ensemble-style extraction principle widely adopted in recent LLM-based CTI studies \cite{wang2023selfconsistency}, all IOC candidates identified by any model are retained to mitigate individual model bias and maximize recall.
Finally, we perform multiple high-temperature inference passes to encourage output diversity, allowing the models to surface less explicit but contextually relevant IOCs, consistent with sampling strategies proposed in \cite{fieblinger2024actionablecyberthreatintelligence}.
Together, these techniques form a practical and reproducible pipeline that provides sufficient IOC coverage for downstream regex generation and evaluation.

We analyzed over \textit{230,000 sentences} and extracted more than \textit{63,000 IOC candidates} as inputs for subsequent regex generation experiments. As shown in Table \ref{ioc_extrated}, 2,302 were registry keys, 12,195 were file paths, and 10,286 were command line arguments—representing the three regex-applicable IOC categories emphasized in Section~\ref{high_value_ioc}.
The remaining 38,820 entries fell into the “Other” category, which includes lower-value indicators such as IP addresses, domain names, hash values, and miscellaneous textual artifacts that are less suitable for regex-based operationalization.

\vspace*{-2mm}
\subsubsection{Ground Truth Strings Dataset for Regex Validation}
While CTI reports provide an excellent input data source for IOC extraction, validating our system's processing results requires ground-truth information that is independent of these reports. However, currently available public datasets cannot provide the necessary ground-truth information for comprehensive validation.  
For example, \cite{scott_freitas_jovan_kalajdjieski_amir_gharib_rob_mccann_2024} encrypts its content due to privacy concerns, and \cite{zhu2023loghublargecollectionlog} does not contain sufficiently rich attack data.

To validate whether the regexes generated by our system can be effectively applied in real-world environments, we utilized data from the MITRE ATT\&CK Evaluation\cite{mie}. The MITRE ATT\&CK Evaluation is an independent and rigorous assessment designed to test whether cybersecurity vendors' products can detect and report various types of attacks. A total of 19 well-known cybersecurity vendors participated, including Bitdefender, Palo Alto Networks, Trend Micro, among others. In this evaluation, 10 well-known attack scenarios (such as APT29, LockBit, etc.) are replicated. In each scenario, there are multiple tactics ranging from Initial Access to Exfiltration, and each tactic may be executed using various techniques and multiple procedures. These vendors provide SIEM screenshots as evidence that they have successfully detected the execution of relevant TTPs. These screenshots contain numerous real-world file paths, registry keys, and command line arguments.
We manually collected all the vendor screenshots and extracted the strings that related to file paths, registry keys, and command line arguments from each one. In total, we collected over 2400 strings across ten attack scenarios as ground-truth, with each attack scenario being a separate dataset. Table \ref{tab:occurrences} shows the number of ground-truth strings in each dataset, as well as how many of those strings are file paths, registry keys, and command line arguments.
\begin{table}[tp]
  \centering
  \small
  \renewcommand{\arraystretch}{1.2}
  \begin{tabular}{lcccc}
    \toprule
    \textbf{Dataset}
      & \makecell[c]{\scriptsize\textbf{File}\\\scriptsize\textbf{path}}
      & \makecell[c]{\scriptsize\textbf{Registry}\\\scriptsize\textbf{key}}
      & \makecell[c]{\scriptsize\textbf{CLI}\\\scriptsize\textbf{args}}
      & \makecell[c]{\scriptsize\textbf{Total}} \\
    \midrule
    Enterprise2024\_LockBit    & 150 & 15  &  42  & 207 \\
    Enterprise2024\_CL0P       &  59 &  5  &  31  &  95 \\
    Turla\_Carbon              & 200 & 12  &  94  & 306 \\
    Turla\_Snake               & 235 &  1  & 131  & 367 \\
    Wizard\_Spider             & 162 &  6  &  91  & 259 \\
    Wizard\_Sandworm           & 102 &  3  &  54  & 159 \\
    APT29\_scenario1           & 150 & 20  &  91  & 261 \\
    APT29\_scenario2           &  50 & 19  & 155  & 224 \\
    FIN7                       &  204 & 5  & 72  & 281 \\
    Carbanak                   &  187 & 16  & 53  & 256 \\
    \bottomrule
  \end{tabular}
  \caption{Ground truth strings count by threat group and scenario}
  \label{tab:occurrences}
  \vspace{-4mm}
\end{table}

\vspace*{-2mm}
\subsection{RQ1: Can IOCRegex-gen generate regular expressions in a scalable way} 
\vspace*{-2mm}
We examine to what extent the regexes generated from the IOCs extracted from 3,156 CTI reports can cover the ground-truth strings in the MITRE ATT\&CK Evaluation datasets, and each extracted IOC is guaranteed to have a corresponding generated regex. Since this study investigates whether our system can generate regexes that accurately retrieve relevant strings from large volumes of log data, we focus specifically on evaluating the accuracy of the regexes themselves.
\begin{table*}[htbp]
  \begin{center}
  \small
  \setlength{\tabcolsep}{4pt}
  \renewcommand{\arraystretch}{1.0}

  \begin{tabular}{@{} l
                     c
                     c
                     c
                     c
                     c @{}}
    \toprule
    \textbf{Dataset} 
      & \makecell{\textbf{Total} \\ \textbf{ground-truth} \\ \textbf{strings}}
      & \makecell{\textbf{\# of Matched} \\ \textbf{Strings}} 
      & \makecell{\textbf{\# of Unmatched} \\ \textbf{(Total/Paths/RegKeys/}\\\textbf{CmdLines})} 
      & \textbf{Hit Rate (\%)} 
      & \textbf{Avg.\ FP Rate (\%)} \\
    \midrule
    Enterprise2024\_LockBit  
      & 207 & 207 
      & \makecell[c]{0\,/\,0\,/\,0\,/\,0}   
      & 100.0 & 0.0 \\

    Enterprise2024\_CL0P     
      &  95 &  94 
      & \makecell[c]{1\,/\,0\,/\,0\,/\,1}   
      &  98.9 & 0.0 \\

    Turla\_Carbon            
      & 306 & 304 
      & \makecell[c]{2\,/\,0\,/\,0\,/\,2}   
      &  99.3 & 1.5 \\

    Turla\_Snake             
      & 367 & 367 
      & \makecell[c]{0\,/\,0\,/\,0\,/\,0}   
      & 100.0 & 1.2 \\

    Wizard\_Spider            
      & 259 & 254 
      & \makecell[c]{5\,/\,1\,/\,0\,/\,4}   
      &  98.1 & 0.5 \\

    Wizard\_Sandworm         
      & 159 & 157 
      & \makecell[c]{2\,/\,1\,/\,0\,/\,1}   
      &  98.7 & 1.3 \\

    APT29\_scenario1         
      & 261 & 255 
      & \makecell[c]{6\,/\,0\,/\,0\,/\,6}   
      &  97.7 & 0.9 \\

    APT29\_scenario2         
      & 224 & 224 
      & \makecell[c]{0\,/\,0\,/\,0\,/\,0}   
      & 100.0 & 0.9 \\

    FIN7         
      & 281 & 278 
      & \makecell[c]{3\,/\,1\,/\,1\,/\,1}   
      &  99.0 & 1.4 \\

    Carbanak        
      & 256 & 254 
      & \makecell[c]{2\,/\,1\,/\,0\,/\,1}   
      &  99.3 & 0.3 \\
    \bottomrule
  \end{tabular}

  \caption{Experimental results of whether regex generation was successful (RQ1) and the false positive rate of the generated regex (RQ2).}
  \label{tab:performance_metrics_reordered_linebreak_fixed}
  \vspace{-4mm}
  \end{center}
\end{table*}

In Table \ref{tab:performance_metrics_reordered_linebreak_fixed}, the ``Number of matched strings" refers to the number of ground-truth strings in each dataset that can be matched by the generated regexes. 
The ``Hit Rate" represents the proportion of these matchable ground-truth strings relative to the total. 
The results demonstrate that regexes generated by our system achieve a 100\% hit rate on ground-truth strings in the Enterprise2024\_LockBit, Turla\_Snake, and APT29\_Scenario2 datasets. In the remaining datasets, although not all ground-truth strings can be fully matched, the lowest hit rate remains 97.7\%. 
Notably, none of the ground-truth strings were derived from any of the 3,156 CTI reports; instead, they originate from an independent source—the MITRE ATT\&CK cybersecurity product evaluation results.

The ``Number of Unmatched Strings" represents the total number of ground‐truth strings in the dataset that were not matched, and also specifies how many of those unmatched strings belong to file paths (Paths), registry keys (RegKeys), or command line arguments (CmdLines).
And we can observe that most of the missed ground‐truth strings fall within the ``command line argument" category. Our analysis identifies two key characteristics shared by these unmatched strings.

First, these cases commonly involve attacker-created executables—that is, custom executables dropped by adversaries during intrusion campaigns, rather than built-in Windows utilities.
For instance, the ground-truth string
\texttt{C:\textbackslash users\textbackslash kmitnick.hospitality\textbackslash appdata\textbackslash loc\\al\textbackslash adb156.exe //b //e:jscript sql-rat.js} in APT29\_Scenario1 represents an invocation of such a malicious payload.
Because these executables are neither part of the standard system toolset nor previously seen in our graph database, the agent lacks sufficient contextual knowledge to model their associated command-line parameters—thus failing to generate matching regexes for this specific IOC type.
Nevertheless, these attacker-created executables usually exhibit distinctive and traceable path patterns, such as anomalous directory locations (e.g., user profile or temporary directories) or randomized filenames.
Even though the agent cannot always reconstruct the full command-line expression, it can still produce accurate regexes that capture the file path portion (e.g.,
\texttt{C:\textbackslash users\textbackslash kmitnick.hospitality\textbackslash appdata\textbackslash loc\\al\textbackslash adb156.exe}), which remains useful for downstream detection.

Second, we discovered that many cmdlets actually used in real-world attacks are not documented in CTI reports. A cmdlet is a type of PowerShell command specifically designed to perform a particular task, such as accessing the file system, managing the registry, or querying the list of services. Attackers often leverage cmdlets to carry out their malicious objectives. For example, in the Wizard\_Spider case, the attacker uses the \texttt{select-string} cmdlet to search for text in strings and files, yet most CTI reports fail to document this usage. Consequently, our system cannot leverage these IOCs as intermediate results to generate the corresponding regexes. However, since we maintain a comprehensive inventory of Windows commands and PowerShell commands (including cmdlets), if a CTI report does include IOCs associated with any of those commands, our system can still use those IOCs to generate the corresponding regexes.
\vspace*{-2mm}
\subsection{RQ2: False Positives of the Generated Regular Expressions} 
\vspace*{-2mm}
Some generated regexes may incorrectly match ground-truth strings they should not target. Each such incorrect match constitutes a false positive. 
We determine false positives by comparing the capture groups in each generated regex against those in its matched ground-truth string.
Specifically, a match is considered a false positive if the regex successfully matches a ground-truth string whose capture groups differ from those of the IOC used to generate the regex. For example, the IOC \texttt{C:\textbackslash Windows\textbackslash System32\textbackslash certutil.exe} has capture groups \texttt{Windows} and \texttt{System32}. The corresponding regex correctly matches a ground truth string such as \texttt{c:\textbackslash windows\textbackslash system32\textbackslash pscp.exe}, which shares identical capture groups with the original IOC. Conversely, if it matches a ground truth string such as \texttt{c:\textbackslash users\textbackslash pam\textbackslash desktop\textbackslash rcs.3aka3.doc}, whose capture groups differ, the match is considered a false positive. To systematically evaluate false positives, we first need to determine the capture group for each ground-truth string. Following the definitions of ``capture group" and ``non-capture group" mentioned in Section \ref{motivation1}, we annotate the capture group for every ground-truth string.
For each generated regex, we check which ground-truth strings it matches and count those whose capture groups differ from the regex as false positives. Then calculate the proportion of these IOCs as the false positive rate, and finally compute, for each dataset, the average false positive rate of regexes for IOCs that match ground truth strings.

We formalize this approach formally. Let $D$ represent the set of all ground-truth strings in the current dataset, and $G(s)$ denote the capture group for any string, $s\in D$. Let $\mathrm{IOC}_k$ denote the source IOC from which $k$-th regular expression $R_k$ is generated. We then define $G_k = G(\mathrm{IOC}_k)$ as the capture groups in that IOC. Define the set of ground‐truth strings that $R_k(s)$ can match as $M_k \;=\; \bigl\{\,s \in D : R_k(s)\text{ matches}\bigr\}$, where $N_k \;=\; \lvert M_k\rvert$ represents the total number of match strings.
Using $\mathrm{FP}_k \;=\; \bigl|\{\,s \in M_k : G(s) \neq G_k\}\bigr|$, we can calculate how many ground-truth strings that $R_k$ should not match but it actually does match (false positives).
The false positive rate for regex $R_k$ is then:
\vspace{-3mm}
\medskip
\begin{equation}
  \label{eq:modified_FPR}
  \mathrm{FPR}\bigl(R_k\bigr)
  \;=\; 
  \frac{\mathrm{FP}_k}{N_k}
  \;=\; 
  \frac{\bigl|\{\,s \in M_k : G(s) \neq G_k\}\bigr|}{\,\lvert M_k\rvert\,}.
\end{equation}

To obtain the overall performance metric for each dataset, we compute the average false-positive rate across all regexes that successfully match at least one ground-truth string:
\vspace{-4mm}
\medskip
\begin{equation}
  \overline{\mathrm{FPR}}
  \;=\; 
  \frac{1}{M}\sum_{\substack{k=1 \\ N_k \ge 1}}^{K} 
  \mathrm{FPR}\bigl(R_k\bigr)
\end{equation}
where $M$ represents the number of regexes that match ground-truth strings in the dataset.

Our evaluation demonstrates that all false-positive rates remain 1.5\% or lower, as shown in Table~\ref{tab:performance_metrics_reordered_linebreak_fixed}. Analysis of the regexes with relatively higher false-positive rates reveals a consistent pattern: such cases occur when attackers deliberately place malicious executables in non-native Windows directories. For instance, paths like \texttt{c:\textbackslash dumps\textbackslash microsoft.activedirectory.webser\\vices.exe.4652.dmp.sk12uyqzk} do not conform to typical Windows directory structures, causing regexes targeting standard system folders to miss them unless the matching patterns are broadened. However, these false positives are readily identifiable and resolvable in real SOC workflows. The matched paths usually deviate from legitimate Windows directory conventions—appearing in user-created folders, temporary dump locations, or application-specific cache directories—making they could be immediately distinguishable by a set of heuristic rules. During log triage, analysts can use these rules to filter such entries in seconds by verifying directory provenance or applying simple whitelists of common benign paths.


\subsection{RQ3: Over-Generalization Assessment}

In RQ3, we investigate whether the generated regexes exhibit over-generalization. We demonstrate that the regexes produced by our system meaningfully incorporate capture-group information rather than relying on overly simplistic patterns such as \texttt{(?i).*cmd.*} or \texttt{(?i).*HKCU.*}.

\begin{figure}[!h]
\centering
\includegraphics[width=\linewidth]{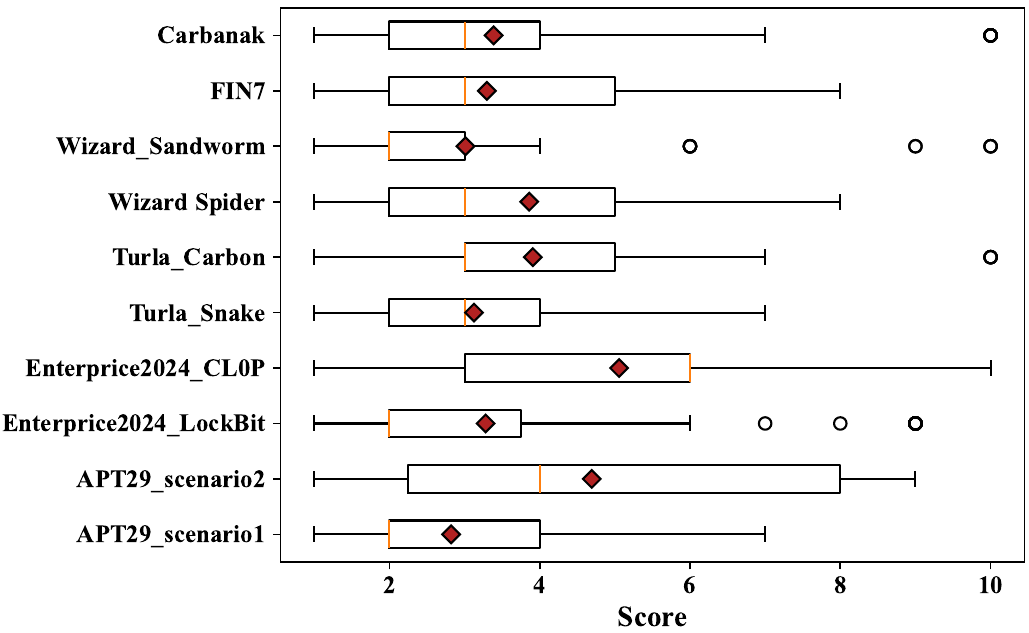}
\caption{The scores of the regular expressions for each dataset.}
\label{fig:scores of the regexes}
\vspace{-3mm}
\end{figure}
To assess the complexity and specificity of our generated regexes, we applied the Regular Expression Grading method described in Section \ref{Regular expression Grading} to score each regex that successfully matched a ground-truth string.
Figure \ref{fig:scores of the regexes} presents the distribution of regex scores across all datasets. The analysis reveals several key findings that demonstrate our system is capable of generating complex regexes.

\noindent\textbf{Overall Complexity Assessment: } The diamond markers indicate mean scores, with most exceeding 3 points. This indicates that, on average, each regex incorporates at least three capture-group strings, demonstrating substantial structural complexity beyond simple pattern matching.

\noindent\textbf{Distribution Characteristics:} The interquartile range shows that, for most datasets, the regexes score between 2 and 6 points, comprising 50\% of all regexes that match the ground-truth strings. This distribution indicates consistent generation of moderately to highly complex patterns rather than overly simplistic expressions.

\noindent\textbf{Dataset-Specific Observations:} Notably, in the APT29\_Scenario2 dataset, the third quartile reaches 8 points, indicating exceptionally complex regexes. This finding aligns with the data in Table \ref{tab:occurrences}, which shows that approximately 70\% of ground-truth strings in APT29 Scenario2 are command line arguments. The structural complexity inherent in command line arguments necessitates more elaborate regexes to capture their diverse patterns.

\vspace*{-2mm}
\subsection{RQ4: Under-Generalization Assessment}
\vspace*{-2mm}
\begin{figure}[!t]
\centering
\includegraphics[width=\linewidth]{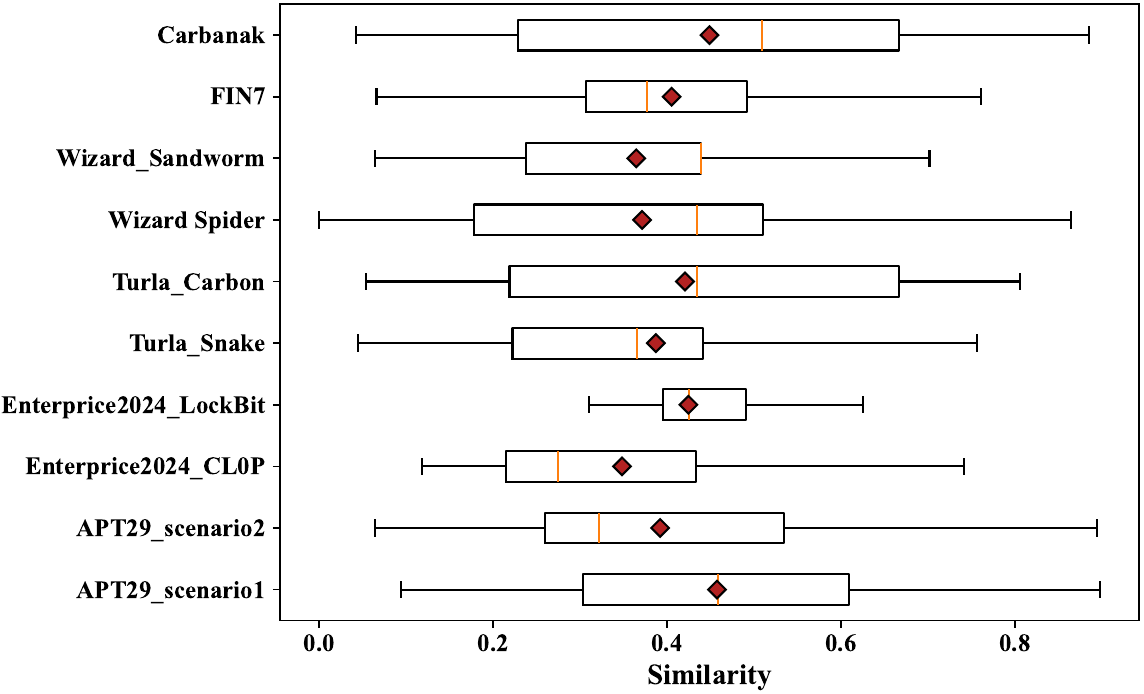}
\caption{Similarity comparison between regular expressions and the original IOCs}
\label{Similarity comparison}
\vspace{-5mm}
\end{figure}

In RQ4, we investigate whether the generated regexes are overly similar to the original IOCs. If regexes are too narrowly tailored to their source IOCs, the set of strings they can match becomes overly restrictive, causing these expressions to fail when matching target strings in different system contexts or machine environments. We use the Levenshtein distance metric \cite{10.5555/1822502} to quantify the similarity between generated regexes (specifically, those that successfully match the respective ground-truth strings) and their source IOCs. 
Figure \ref{Similarity comparison} presents the similarity distribution across all datasets, where values increase from left to right. The analysis reveals several important patterns regarding the relationship between generated regexes and their source IOCs.

\noindent\textbf{Overall Similarity Characteristics: } The average similarity hovers around 0.4, indicating a moderate level of similarity that suggests the generated regexes maintain some connection to their source IOCs while incorporating sufficient variation for broader applicability.

\noindent\textbf{Distribution Spread:} Both the first and third quartiles lie predominantly between 0.2 and 0.6, demonstrating that the generated regexes generally differ substantially from the original IOCs. This distribution indicates that our system successfully generates expressions that are neither too similar (overly restrictive) nor too dissimilar (potentially inaccurate) to their source material.

\noindent\textbf{High Similarity Cases:} Some regexes exhibit exceptionally high similarity to the original IOCs. Our analysis reveals that this phenomenon tends to occur predominantly in the file-path and registry-key categories. This pattern emerges because in these IOC types, a substantial portion of the substring is captured by capture groups. Since our system preserves these capture groups when constructing regexes, the resulting expressions remain very close to the originals.
For example, in the IOC \texttt{ProgramData\textbackslash Microsoft\textbackslash Windows\textbackslash StartMenu\textbackslash P\\rograms\textbackslash StartUp}, each directory name within the path serves as an individual capture group, resulting in the generated regex \texttt{(?i).*ProgramData\textbackslash\textbackslash Microsoft\textbackslash\textbackslash Windows\textbackslash\textbackslash Sta\\rtMenu\textbackslash\textbackslash Programs\textbackslash\textbackslash StartUp.*}.
\vspace*{-2mm}
\subsection{RQ5: Ablation Study}
\vspace*{-2mm}
\begin{table*}[ht]
  \centering
  \resizebox{0.8\textwidth}{!}{%
    \begin{tabular}{lcccccc}
      \toprule
      Dataset
        & -CR Hit (\%)
        & -CR Avg. FP (\%)
        & C-R Hit (\%)
        & C-R Avg. FP (\%)
        & -C-R Hit (\%)
        & -C-R Avg. FP (\%) \\
      \midrule
      Turla\_Carbon
        & 82.6 & 18.4 & 73.5 & 57.7 & 74.5 & 56.8 \\
      Turla\_Snake
        & 80.9 & 24.2 & 76.0 & 24.5 & 80.4 & 25.9 \\
      Wizard\_Spider
        & 66.4 & 19.4 & 60.4 & 36.2 & 61.0 & 39.3 \\
      APT29\_scenario1
        & 77.7 & 12.7 & 66.2 & 26.5 & 65.5 & 18.0 \\
      \bottomrule
    \end{tabular}%
  }
  \caption{Ablation study of regular expression generation}
  \label{tab:ablation}
\end{table*}
To evaluate the contribution of individual components to our system's performance, we conducted ablation experiments on four datasets. These datasets contain a large number of ground-truth strings and maintain a balanced distribution across the file path, registry key, and command line argument categories. We believe they are sufficiently representative for ablation experiments. 

For each dataset, we systematically disabled two key components of our system and created three experimental settings. The first ``-CR” omits the Capture Group Finding step but retains the Reasoning-based Regex Generation step for generating regexes. The second ``C-R” includes the Capture Group Finding step but removes the Reasoning-based Regex Generation step when generating regexes. The third, ``-C-R,” indicates that both the Capture Group Finding step and the Reasoning-based Regex Generation step are disabled simultaneously. When comparing Table \ref{tab:performance_metrics_reordered_linebreak_fixed} with Table \ref{tab:ablation}, regardless of which component is disabled, both the hit rate and false‐positive rate under ablation settings fall substantially short of the baseline results: the hit rate drops by over 16.7\% and the average false‐positive rate rises by over 23.3\%. This significant performance degradation demonstrates that each component plays an essential role in generating regexes capable of accurately capturing the ground‐truth strings.

From Table \ref{tab:ablation}, we observe that, in most cases, the hit rate of our system using the Reasoning-based Regex Generation step but without the Capture Group Finding step is slightly better than that of the system without the Reasoning-based Regex Generation step but with the Capture Group Finding step, as well as better than the system with neither component. Notably, in the Turla Carbon dataset experiment, the regexes generated by the system with the Reasoning-based Regex Generation step had a hit rate more than 8\% higher than those generated by the other two without it. This performance difference is attributed to the Reasoning-based Regex Generation’s role in generating a more comprehensive set of valid regexes. Without the Reasoning-based Regex Generation step, a valid regex cannot be guaranteed for each IOC, resulting in an inability to match every ground-truth string and a lowered overall hit rate.
Moreover, when the Reasoning-based Regex Generation step is absent, the reduced number of valid expressions correlates with simpler structural patterns in the generated regexes. These simpler structures tend to be more general in nature, leading to increased false-positive rates as they inadvertently match strings beyond their intended scope. This observation is also reflected in the Turla Carbon dataset experiment. Compared to the system with the Reasoning-based Regex Generation step but without the Capture Group Finding step, the system without the Reasoning-based Regex Generation step showed a 38.4\% increase in average false positive rate. 
\vspace*{-2mm}
\subsection{Cross-Model Consistency of Regex Generation}
\vspace*{-2mm}
To further examine whether our system's behavior is model-dependent or stable across different foundation models, we compare the structural similarity of regexes generated by GPT-4o, LLaMA3, and DeepSeek-V3. We adopt a feature-based syntactic comparison approach: we extract core structural elements—including grouping constructs, character classes, wildcard tokens, anchors, quantifiers, alternation operators, and escape symbols—and encode them into structural feature vectors. Cosine similarity is then applied to quantify the syntactic resemblance of model outputs. The results show high consistency across models: the mean similarity between GPT-4o and LLaMA3 is 0.980, and between GPT-4o and DeepSeek-V3 is 0.974.

\vspace*{-2mm}\section{Related Works}
\label{related_work}
\vspace*{-2mm}
\subsection{Extracting Information From CTI}
\vspace*{-2mm}
In recent years, advances in NLP have significantly improved the extraction of crucial information from CTI. TTPDrill \cite{Husari2017} applies NLP to automatically identify attack patterns (TTPs) from CTI data, while TINKER \cite{Rastogi} leverages pre-trained NER and RE models to build structured knowledge graphs for large-scale threat intelligence sharing. \cite{Gao} proposes an unsupervised NLP pipeline for extracting structured threat behaviors from unstructured OSCTI texts, along with a specialized query language for efficient querying. With the rise of LLMs, researchers have begun using them to extract richer and more complex threat intelligence \cite{hu2024llm, huang2024ctikg, liu2023constructing, cuong2025towards, fieblinger2024actionable}. For instance, \cite{hu2024llm} employs LLMs to extract and annotate CTI entities and TTPs, \cite{huang2024ctikg} constructs security knowledge graphs using multi-agent LLMs with dual-memory mechanisms, and \cite{liu2023constructing} guides ChatGPT to act as a “cybersecurity expert” for ontology-constrained metadata extraction.

\textbf{LLM-based Rule Generation.} Furthermore, some papers go beyond merely extracting information from CTI and instead leverage that information to generate rules for detecting attacks. For example, in \cite{Xu2024}, the authors leverage the in-context learning capabilities of LLMs to perform TTP classification. In this approach, the authors combined an external database with LLM-based extraction of attack-level threat intelligence to improve the model's discriminative capabilities. According to their released dataset of 135 generated rules, most correspond to indicators such as IP addresses and domain names, while only 7 rules involve any form of regex-based pattern matching. Despite advances in automatic rule creation, automating the regex portions of detection rules—which are crucial for precise pattern matching—remains underexplored. Additionally, \cite{Wang2026RulePilot} proposed an LLM-powered agent that autonomously generates executable detection rules for SIEM platforms. Their approach leverages semantic reasoning, intermediate representations, and iterative refinement to synthesize complex behavioral rules that can be directly executed within SIEM environments. However, they focus primarily on high-level rule logic and do not explicitly address the automation of low-level regex construction for precise indicator matching. Moreover, in \cite{Schwartz2024}, the authors proposed a novel framework to automatically generate cloud-oriented detection rule candidates from cloud-based CTI. They designed a three-step process to complete the entire workflow, comprising preprocessing, TTP extraction, and rule generation, and employed LLMs to carry out both the TTP extraction and the rule generation stages. However, analysis of the public Splunk detection rules \cite{Splunk_Security_Content} shows that such cloud-oriented rules constitute only a small fraction of real-world SIEM rules, and these rules rarely incorporate regular expressions as matching mechanisms. 

\vspace*{-2mm}
\subsection{Regular Expression Generation}
\vspace*{-2mm}

Regex generation has long been an active research topic. Early studies primarily relied on heuristic approaches. For instance, \cite{10.1145/2576768.2598333} employed Genetic Programming (GP) to generate minimal regexes that distinguish between positive and negative examples, while \cite{10.1145/2993231.2993232} extended this idea by using GP as a solver within an active learning framework to evolve high-performing regexes for text extraction. Similarly, \cite{uzun2020regular} proposed a heuristic method that derives regexes directly from CSS selectors, reducing the computational overhead of DOM-based parsing. With the rise of deep learning, later works explored neural models for translating natural language into regex patterns, improving both accuracy and applicability. For example, \cite{9401951} proposed NLP-based synthesis and regex repair algorithms, \cite{ye2020sketch} introduced a semantic parsing framework that converts natural language into regex sketches for program synthesis, and \cite{locascio2016neuralgenerationregularexpressions} utilized a two-stage Long Short-Term Memory (LSTM) model for regex generation. More recently, researchers have begun leveraging large language models (LLMs) for this task. In \cite{10.1145/3643916.3644424}, LLMs are prompted to generate regexes from natural language descriptions, while \cite{zelinaexamples} applies a feedback-based LLM approach to iteratively refine regexes for clinical text extraction. However, none of these approaches effectively addresses the unique challenges of regex generation for IOCs. The adversarial and highly variable nature of IOCs, coupled with the scarcity of diverse positive and negative examples, limits the performance of existing methods when applied to security-oriented regex generation.

\vspace*{-2mm}\section{Conclusion}
\vspace*{-2mm}
In this paper, we introduce a novel LLM-based system that automates the translation from extracted IOCs to deployable regexes. Using our proposed framework, the system automatically identifies the essential components within an IOC that are necessary for generating regular expressions, significantly enhancing the accuracy of the resulting regex patterns. Our experimental results show that the regex patterns generated by our system match the ground truth strings with exceptional accuracy, demonstrating that our system can perform effectively in real-world scenarios.
\section*{Ethics considerations}
None.



\bibliographystyle{IEEEtran}
\bibliography{references.bib}
%



\end{document}